\documentclass[pre,a4paper,final,showkeys]{revtex4}
\usepackage[]{graphicx}
\usepackage{amsmath}
\usepackage{psfrag}

\newcommand{\bq}{\begin{equation}}
\newcommand{\eq}{\end{equation}}
\newcommand{\ba}{\begin{eqnarray}}
\newcommand{\ea}{\end{eqnarray}}

\begin{document} 

\title{Free zero-range processes on networks} 

\author{L. Bogacz$^{a}$}
\author{Z. Burda$^{b,c}$}
\author{W. Janke$^{d,e}$}
\author{B. Waclaw$^{b,d}$}
\affiliation{
$^{a}$Department of Information Technologies, \mbox{Faculty of Physics, Astronomy and Applied Informatics}, Jagellonian University, Reymonta 4, 30-059 Krak\'ow, Poland \\
$^{b}$Marian Smoluchowski Institute of Physics, Jagellonian University, Reymonta 4, 30-059 Krak\'ow, Poland \\
$^{c}$Mark Kac Complex Systems Research Centre, Jagellonian University, Krak\'ow, Poland \\
$^{d}$Institut f\"ur Theoretische Physik, Universit\"at Leipzig, Postfach 100\,920, 04009 Leipzig, Germany \\
$^{e}$Centre for Theoretical Sciences (NTZ), Universit\"at Leipzig, Germany
}
 
\begin{abstract}
A free zero-range process (FRZP) is a simple stochastic process describing the
dynamics of a gas of particles hopping between neighboring nodes
of a network. We discuss three different cases of increasing 
complexity: (a) FZRP on a rigid geometry where the network is fixed during the process,
(b) FZRP on a random graph chosen from a given ensemble of networks, %that is all physical quantities are averaged over graphs from the ensemble,
(c) FZRP on a dynamical network whose topology continuously
changes during the process in a way which depends on
the current distribution of particles.
The case (a) provides a very simple 
realization of the phenomenon of condensation which manifests 
as the appearance of a condensate of particles on the node 
with maximal degree. A particularly interesting example is the condensation on scale-free networks. Here we will model 
it by introducing a single-site inhomogeneity to a $k$-regular
network. This simplified situation can be easily treated analytically 
and, on the other hand, shows quantitatively the same behavior 
as in the case of scale-free networks. The case (b) is very interesting 
since the averaging over typical ensembles of graphs acts as a kind 
of homogenization of the system which makes all nodes
identical from the point of view of the FZRP. 
In effect, the partition function of the steady state becomes
invariant with respect to the permutations of the particle occupation
numbers. This type of symmetric systems has been intensively
studied in the literature. In particular, they undergo a phase transition to the condensed phase, 
which is caused by a mechanism of spontaneous symmetry breaking. 
In the case (c), the distribution of particles and the dynamics of network
are coupled to each other. 
The strength of this coupling depends on the ratio of two time scales: for changes of the topology and
of the FZRP.
We will discuss a specific example of that type
of interaction and show that it leads to an interesting phase diagram. 
The case (b) mentioned
above can be viewed as a limiting case where the typical time scale of
topology fluctuations is much larger than that of the FZRP.

\end{abstract}


\maketitle 

%>>>> Include a list of keywords after the abstract 

\section{INTRODUCTION}
\label{sect:intro}  
In the last decade great progress has been achieved in our
understanding of the structure and topology of complex networks.
Complex networks became a sui generis branch of research on 
the interface between physics, probability and graph theory,
geometry, computer science and many other disciplines
and they attracted a lot of researchers.
Probably the main interest in networks comes 
from the fact that in many applications they can be viewed as a skeleton of complex systems,
along which various signals and information propagates between various
components of the system. The dynamics of this propagation 
is responsible for the functionality of complex systems and depends on the structure of the underlying network. 
In some situations the structure can flexibly adjust during the
evolution of the system to optimize the propagation and functionality,
leading to a feedback of the dynamics on the structure of a network. In the most general case, the
dynamics of degrees of freedom propagating on the complex network 
and the dynamics of the network topology are intertwined.
%They interact together to realize a common 
%goal of the optimization of functionality of the system.

In this paper we shall discuss a model describing the dynamics
of some particles propagating through the network. We shall analyze three cases of gradually increasing complexity:
starting from the model
on a given network, through the model averaged over networks, and ending with a model where the dynamics of particles and of
network topology are coupled to each other. It is clear that to
realize this plan one has to consider a relatively simple model
to be able to make some predictions analytically and, on the other
hand, complicated enough to reveal some interesting properties. 
A very good candidate is a gas of identical particles (or balls) hopping 
between neighboring nodes of the network. At each
time step, one particle hops from every non-empty node,
to one of its neighboring nodes which is chosen at random. 
It is a particular example of a more general process called
zero-range process (ZRP), where the hopping rate only depends on the
number of particles at the departure node. Since in our case
the hopping rate is identical for all nodes and particles 
hop unconditionally we call this process free zero-range process (FZRP).
Despite its simplicity the model has many interesting features. 
The system of particles has a steady state and it 
undergoes a phase transition from a liquid state to the state where a
condensate of particles is formed on a single node.
The statics of the steady state is so simple that
it can be analyzed analytically. Also many aspects of its dynamics
can be well understood by analytical methods. 
It is probably the simplest model with non-trivial behavior
where many aspects of the underlying network structure and its
interaction with the degrees of freedom defined on the network 
become pronounced and play an essential role.

The paper is organized as follows. 
We first recall the main results on the FZRP on a static network, 
then we discuss FZRPs averaged over random networks from a given ensemble. 
We pay a special attention to the effect of homogenization
and discuss the relation between the node degree distribution
of the random network and the particle distribution of the steady
state of the FZRP, defined on this network. Finally, we 
discuss a simple model of FZRP on a dynamical random network
with an explicit coupling between the dynamics of particles
and of network topology. At the end of paper we shortly summarize our main results.

\section{STATICS AND DYNAMICS IN FREE ZRP} 
\label{sect:statics}
Consider a system of $M$ identical undistinguishable particles 
and a graph with $N$ nodes. The dynamics is driven by
a stochastic process defined in such a way that at
the given moment of time one particle jumps from each non-empty 
node to a randomly chosen neighboring node on the network.
In general one can think of a synchronous dynamics but in practice
when one wants to realize the process by a 
simple Monte Carlo computer simulation, 
one picks up nodes at random, one by one, and if the given node is not empty, 
one moves a particle to a neighboring node chosen with
probability $1/k_i$. Here $k_i$ is the number of neighbors 
of the node $i$, called its degree. On average,
the outflow of particles along each link is equal to $1/k_i$
if there is at least one ball at that node.
The full state of the system is given by the distribution 
of particles $\{m_1,\ldots,m_N\}$, where $m_i$ is the number
of particles at the node $i$. The total number of particles 
$M=m_1+\ldots+m_N$ is conserved during the process. 
The model described above is a special case of the 
zero-range process (ZRP)\cite{evans2,godreviews, cc1, cc5} 
with no point-interaction between balls. 

If the network is connected, the FZRP has a unique steady state 
\cite{evans}. Although it is a non-equilibrium state, one can formally 
write for it a partition function $Z(N,M,\vec{k})$ 
which for the given graph depends on the sequence of the node degrees 
$\vec{k}=\{k_1,...,k_N\}$ but not on other details of the graph topology:
\bq
Z(N,M,\vec{k}) = 
\sum_{m_1=0}^M\cdots\sum_{m_N=0}^M \delta_{m_1+\ldots +m_N, M}
\prod_{i=1}^N p_i(m_i),	\label{part2}
\eq
where $p_i(m)=k_i^{m_i}$.
The only constraint that prevents  $Z(N,M,\vec{k})$ 
from a full factorization is the discrete delta function 
reflecting the conservation of particles.
The partition function (\ref{part2}) has exactly the same form 
as the balls-in-boxes model \cite{bbj,bbj2}.
Since, beside the degree sequence, other topological features
of the network are irrelevant for the steady state,
%and actually also for the most rude characteristics of the dynamics
one can forget about the network and think only of balls hopping between boxes having statistical weights $p_i(m)$,
which may be different for different boxes $i$.
The knowledge of the partition function allows one to calculate 
all static quantities of the model and well approximate most of 
the dynamical quantities describing its behavior 
out of the steady state. Similarly as in the ZRP, the most interesting
quantity which allows one to grasp what happens in the system
is the node occupation distribution $\pi_{i,\vec{k}}(m)$
which tells one the probability that the node $i$
has exactly $m$ balls. It can be calculated as 
\bq
\pi_{i,\vec{k}}(m) = \frac{Z(N-1,M-m,\vec{k}_i)}{Z(N,M,\vec{k})} p_i(m), 
\label{piim}
\eq
where $Z(N-1,M-m,\vec{k}_i)$ is the partition function (\ref{part2})
for a graph with $N-1$ nodes, $M-m$ particles
and degrees $\vec{k}_i \equiv \{k_1,\dots,k_{i-1},k_{i+1},\dots,k_N\}$,
where $k_i$ is skipped. If one averages Eq. (\ref{piim}) over nodes
of the network, one obtains the probability that 
a randomly chosen node is occupied by $m$ particles:
\bq
\pi(m) = 
\frac{1}{N} \sum_{i=1}^N \pi_{i,\vec{k}}(m). \label{pigeneral}
\eq
We shall see below, that this global quantity can be used as a signature of condensation.
If all $k_i$'s are the same, as it is for a $k$-regular graph, 
the system is said to be homogeneous. In this case all nodes in the partition function (\ref{part2})
are statistically equivalent 
and the quantities averaged over the nodes
are identical to those obtained separately for any node, for
instance $\pi_i(m)=\pi(m)$ for all $i$'s.
Moreover, the degree $k$ appears in Eq. (\ref{part2})
as a factor $k^M$ which can be pulled out in front of the sum
as a constant and which cancels out in physical quantities 
like the distribution of balls (\ref{pigeneral}). 
A more interesting case is of course when the degrees $k_i$ 
vary from node to node and the system is inhomogeneous.
In this case in order to fully characterize the system
one should indeed independently determine $\pi_i(m)$ 
for every node $i$. 

We are often interested in the behavior of the 
system for a given density of particles $\rho=M/N$
in the thermodynamic limit, that is when the size of the system $N$ 
tends to infinity. What makes the FZRP (or generally the ZRP) 
so interesting is that by increasing the density 
one may trigger off the phenomenon of condensation. % in the system.
Above a certain density of balls $\rho_c$ an extensive number of
particles, proportional to the surplus above the critical
density: $\Delta M=M-\rho_c N$, condenses on a single node --
the one with highest degree. The mechanism underlying
the condensation can be understood by comparing flows
of particles coming in and going out of a node.
The average outflow of particles from a node $i$ does not depend
on its degree, but the inflow does. It is proportional 
to $\sum_j 1/k_j \sim k_i/\left<k_j\right>$ 
where the sum runs over all non-empty neighbors of $i$. 
If $k_i$ is greater than any $k_j$ and balls are uniformly distributed, 
the outflow is smaller than the inflow. Thus the node $i$ attracts 
more and more particles, as long as there are enough balls in the vicinity 
of $i$. When the density of balls in the neighborhood falls below 
some value, the in- and outflows balance each other.
This process leads to a fast local condensation on a few nodes with 
higher degrees. Then, by exchanging particles through the background, 
all partial condensates merge into a single one.

\begin{figure}
\begin{center}
\psfrag{xx}{$m$} \psfrag{yy}{$\pi(m)$}
\includegraphics*[width=14cm]{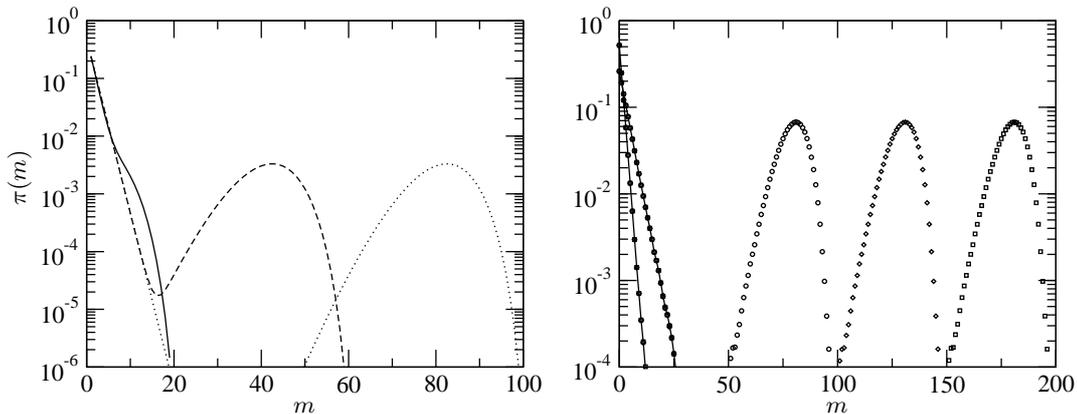}
\end{center}
\caption{\label{inh+BA}Left: theoretical distributions of balls $\pi(m)$ for a single inhomogeneity graph with $k_1=8,k=4,N=20$, for three different $M$: $20$ (solid line), $60$ (dashed), and $100$ (dotted). The critical density is $\rho_c=1$. Peaks are located at $m_*\approx M-\rho_c N$. Right: distributions for a single B-A network of size $N=100,L=198$ and $M=100$ (circles), $150$ (diamonds) and $200$ (squares), from computer simulations. The three peaks show $\pi(m)$ for the node with the highest degree $k_1=27$, while the two left-most lines are for $k_2=20$ (the next largest degree) and $k_3=k_4=13$. An exponential decay is clearly seen in $\pi_{2,3,4}(m)$.}
\end{figure}

A short inspection of the partition function (\ref{part2})
makes it clear that the condensate has the largest chance 
to form on a node with the largest degree, since the statistical
weight of this node $p_i(m) = k_i^m$ favors 
larger values of $m$ stronger than other nodes. Therefore,
while analyzing the static properties, one
should concentrate on this node. 
In order to simplify the discussion we consider first graphs having
all but one regular nodes of degree $k$, and
a single irregular node of degree $k_1>k$.
It is convenient to introduce in this case a parameter 
$\alpha=k/k_1$ which controls the strength of inhomogeneity.
The distribution of balls at the singular node can be 
calculated from the partition function (\ref{part2}) 
with $p_1(m)=k_1^m$ and $p_i(m)=k^m$ for $i>1$. 
The result is \cite{zrp-long}
\bq
\pi_1(m) \propto \alpha^{-m} \binom{M+N-m-2}{M-m}. \label{pi1}
\eq
This distribution has a maximum at $m_*\approx M-\rho_c N = (\rho-\rho_c)N$, 
with a critical density $\rho_c=\alpha/(1-\alpha)$. We see that
above $\rho_c$, an extensive number of particles condense at 
the irregular node. The critical density is larger than zero 
for $\alpha<1$, that is when the degree of the irregular node 
is larger than the degrees of other nodes. When $k_1$ becomes equal to $k$ or
smaller, the critical density becomes infinite and the system never enters the 
condensed phase. One can also show that the occupation 
distribution for any regular node falls exponentially for $\rho>\rho_c$,
\bq
\pi_{\rm reg}(m) \propto \alpha^m , \label{pii}
\eq
which means that the condensate does not appear on regular nodes. 
The average distribution of balls is given by 
$\pi(m)=\left[(N-1)\pi_{\rm reg}(m)+\pi_1(m)\right]/N$.
From Eqs.~(\ref{pi1}) and (\ref{pii}) we see that for a fixed density 
the condensation manifests in $\pi(m)$ as a peak whose position moves linearly with the system size $N$.
The area under the peak is equal to $1/N$ 
since the condensate is located at a single node: one out of $N$.
On the left-hand side of Fig. \ref{inh+BA} we show plots 
of $\pi(m)$ calculated for different densities. One sees 
that the position of the peak moves linearly with $M$.

This simple case of condensation on a graph with
a single irregular node captures well what happens
on general inhomogeneous networks.
This is confirmed by numerical simulations on Barab\'{a}si-Albert (B-A) scale-free graphs \cite{cn}, 
with the degree distribution $\Pi(k)$ falling like $\Pi(k)\sim k^{-3}$. 
The network is built of a small number of high-degree nodes, called hubs, 
to which many nodes with small degrees are linked. From the point of view
of the FZRP the hubs behave similarly as the irregular nodes in the simple
model discussed above. On the right-hand side of Fig.~\ref{inh+BA} we show 
the occupation distribution $\pi(m)$ for one particular B-A network 
with $N=100$ nodes and $L=198$ links, as obtained in Monte Carlo computer simulations. 
The maximal degree is $27$. 
Similarly to the situation for the single inhomogeneity graph, 
the condensation takes place only at the node with highest degree. 
The critical density can be estimated from the position of peaks: 
$\rho_c\approx 0.19$. This value is close to that obtained using
the formula derived for the single inhomogeneity graph: 
$\rho_c=\alpha/(1-\alpha) \approx 0.17$, if $\alpha$ is assumed 
to be $\bar{k}/k_{\rm max}\approx 4/27$. In short, as anticipated
the statics is almost identical to that of the single inhomogeneity 
model.

We shall now shortly discuss the dynamics of the condensation. 
The process of condensate formation from a state where all 
nodes have approximately the same number of balls has been investigated 
in ZRP's \cite{evans, god, grosk, jdn} and can be divided into two steps.
First, the surplus of balls is accumulated on some nodes having 
relatively high degrees in comparison to the rest. Then, these small 
condensates merge and form a single one. The characteristic time-scale for this process depends on the structure on the network, but in general it grows like a power of the system size $N$. 
Once formed, the condensate is not a static object but fluctuates and can sometimes melt down to be indistinguishable from the background. It rebuilds however very quickly. One may ask about the typical life-time $\tau$ of the condensate, that is how much time it takes to fall from $\approx \Delta M$ balls at the condensed node to $\approx \rho_c$, being the average occupation for the ``background nodes''. In Ref.~\cite{god} a mean-field procedure for calculating $\tau$ has been derived. It relies on the assumption that the state of the condensate varies slowly in comparison to the time scale of fluctuations on all 
remaining nodes in the system. This is a good approximation if the network is compact, that is if its diameter grows slowly with $N$, for example like a logarithm, because then balls travel very fast through the graph. Actually in many complex networks one observes such a slow increase of the diameter.
One can therefore distinguish a single ``slow'' variable giving the number of balls at the node where the condensate is located, and apply a mean-field dynamics treating the rest of the system as a steady-state. In this way one obtains a closed formula for the mean time $\tau_{mn}$ it takes to decrease the occupation of the
node in question from $m$ to $n$ particles~\cite{zrp-long,god,phd-thesis}.
It can be in particular used to determine the life-time of the condensate
in the model with a single inhomogeneity~\cite{phd-thesis}.
The full formula is quite complicated, but the life-time can be 
estimated qualitatively with the help of the Arrhenius law~\cite{arh} which
 states that the average life-time is inversely proportional to 
the exponent of the barrier of the effective potential. Since the
effective potential is equal to $-\log \pi_1(m)$ up to a normalization,
the exponent just gives the minimal value of the distribution $\pi_1(m)$. 
Taking into account the proper normalization of Eq.~(\ref{pi1}), 
one gets $\tau\sim (k_1/k)^{\rho N}$. The time thus grows exponentially 
with the system size $N$, while the characteristic time for building 
the condensate from a uniform distribution of particles grows only 
like a power of $N$. This is different from what one observes for
homogeneous systems, where both the times grow like a power, each 
having a different exponent, of the system size \cite{god}.

\section{FREE ZRP ON QUENCHED RANDOM NETWORKS} 
\label{sect:fluct}

In the previous section we have shown results of Monte Carlo simulations 
for a single B-A scale-free network. One can ask what happens if one considers 
FZRP not on a single network, but rather on the whole set of networks. 
We define therefore a statistical ensemble of networks, where each graph has a certain probability of occurrence. All physical quantities, as for instance
the occupation probability, have to be now averaged over this ensemble. 
Because static properties of the ZRP depend only on the sequence of degrees $\vec{k}$, we can reduce the problem to averaging with weights $P(\vec{k})$ giving probabilities of occurrence of networks with a degree sequence $\vec{k}$.
The effective occupation distribution of balls averaged over
the ensemble then reads
\bq
\pi(m) = \frac{1}{N} \sum_i \sum_{k_1,\dots, k_N} P(\vec{k}) \, \pi_{i,\vec{k}}(m),
\eq
where $\pi_{i,\vec{k}}(m)$ is given by Eq. (\ref{piim}) for each individual
sequence $\vec{k}$. It is important to realize that, for typical ensembles of graphs,
the probability distribution $P(\vec{k})$ is invariant with respect 
to a permutation of degrees $k_i$ in the sequence 
$\vec{k}=\{k_1,...,k_N\}$. This is because labeling of nodes plays often only an auxiliary role
and is unphysical. The invariance with respect
to permutations means that each term in the sum 
over $i$ in the last equation is identical and independent of $i$. 
It is therefore sufficient to calculate it for the first node and 
average over the ensemble:
\bq
\pi(m) = \sum_{k_1,\dots, k_N} P(\vec{k}) \, \pi_{1,\vec{k}}(m).
\eq
The system becomes now homogeneous and the distribution of balls
is independent of the node.
The partition function $Z(N,M)$ has now the following form:
\bq
Z(N,M) =\sum_{k_1,\dots, k_N} P(\vec{k}) \, Z(N,M,\vec{k}), 
\label{canon}
\eq
where $Z(N,M,\vec{k})$ is given by Eq.~(\ref{part2}). 
In general, $P(\vec{k})$ could have a complicated form. We shall restrict 
here to ensembles of quenched networks with a product measure sometimes called 
uncorrelated networks. One can find explicit canonical and grand-canonical
realization of such ensembles \cite{homnasz} for which
the probability $P(\vec{k}) = \Pi(k_1)\cdots \Pi(k_N)$ factorizes 
in the limit $N \rightarrow \infty$, where $\Pi(k)$ denotes the probability distribution of the node degrees. 
The difference between the canonical ensemble, having a fixed number of edges,
and grand-canonical one disappears in the thermodynamic limit
if the node degree distribution $\Pi(k)$ 
falls with $k$ faster than any power-law \cite{dorog}.
This factorization partially breaks down for scale-free networks 
which we shall not discuss here. The factorization allows us 
to rewrite the formula for $Z(N,M)$ in the form 
of Eq.~(\ref{part2}) with $p_1(m)=\dots=p_N(m)\equiv \mu(m)$,
\bq
Z(N,M) = 
\sum_{m_1=0}^M\cdots\sum_{m_N=0}^M \delta_{m_1+\ldots + m_N , M}
\prod_{i=1}^N \mu(m_i),	\label{part3}
\eq
where $\mu(m)$ is $m$-th moment of the degree distribution $\Pi(k)$,
\bq
\mu(m) = \sum_{k=1}^\infty \Pi(k) k^m. \label{mom}
\eq
In contrast to Eq. (\ref{part2}) the partition function (\ref{part3})
is invariant under permutations of the ball occupation 
numbers $m_i$. We see that the problem of finding the distribution of balls 
in the ensemble of uncorrelated networks reduces to the ZRP on homogeneous networks, with weights given by Eq.~(\ref{mom}). 
Zero-range processes on homogeneous networks with identical weight functions
for all nodes were intensively investigated in the past. In this paper we shall cite some of the most 
important results and compare them to those on
inhomogeneous graphs described in the previous section. We will also
use them to solve the specific case (\ref{part3}) where the 
weights are given by the moments of the distribution $\Pi(k)$.
It is quite surprising that the degree sequence of the underlying
network entirely determines the distribution of particles and
other properties of the steady state of the FRZP defined on these networks.

The partition function of a homogeneous system 
is in general given by Eq.~(\ref{part2}) with $p_i(m)\equiv p(m)$, where $p(m)$ 
is some arbitrary weight function. The critical properties of the model
depend on the asymptotic behavior of $p(m)$. First, one should notice
that if one rescales the weight as follows: $p(m) \rightarrow e^{a + bm} p(m)$
then the partition function changes by a multiplicative constant: 
$Z(M,N) \rightarrow e^{Na+Mb} Z(M,N)$. Such a constant 
prefactor in the partition function does not change physical quantities. 
Therefore, choosing $p(m) \rightarrow e^{a + bm} p(m)$ one can 
get rid of exponential growth or decay and consider only the remaining
large $m$ asymptotics of $p(m)$. 
%We assume that $p(m)$ cannot grow faster than exponentially. 
The asymptotic behavior can be classified into three groups:
$p(m)$ falls to zero faster than any power of $m$, 
falls like a power $\sim m^{-b}$, or approaches a positive constant.
In the last case the system belongs to the same
universality class as a system of non-interacting 
balls and thus it is equivalent to a random walk of $M$ particles on a 
homogeneous graph. There is no condensation in this case. 
Similarly, there is no condensation
when the weight $p(m)$ decreases with the number of balls $m$
faster than a power-law since there exists an effective repulsive 
force between balls preventing them from occupying a single site. 
Balls tend to distribute on the whole graph and the critical density is $\rho_c=\infty$.
The most interesting case is when 
$p(m)$ falls like a power of $m$: $p(m)\propto m^{-b}$. There are two subcases. In the first one,
$0\le b<2$ and the attraction between
particles is still too weak to form a condensate. In the second and 
the most interesting one, $2<b<\infty$ and a qualitatively different picture emerges. The critical
density is finite in this case. The attraction is strong enough 
to trigger the condensation if the density $\rho$ is larger than
$\rho_c$. At the critical density, the distribution of balls 
$\pi(m)\sim m^{-b}$ falls like a power-law. 
When $\rho$ exceeds $\rho_c$, the distribution has the same shape
as the critical one but it additionally develops a peak
which departs from the `bulk' distribution when $N$
goes to infinity. The area $1/N$ under the peak tells us that the
condensate occupies a single site (one of $N$) almost all time. The site
is chosen randomly from all $N$ nodes by spontaneous symmetry breaking, in contrast to inhomogeneous systems, 
where the symmetry is explicitly broken. Sometimes the condensate melts 
and then rebuilds at another node. The characteristic life-time $\tau$ 
can be estimated \cite{god} as $\tau\propto (\rho-\rho_c)^{b+1} M^b$. 
Because $M=\rho N$, the life-time grows like a power of $N$, contrary 
to the formerly discussed inhomogeneous graphs where it grows exponentially.

\begin{figure}
\begin{center}
\psfrag{xx}{$m$} \psfrag{yy}{$\pi(m)$}
\includegraphics*[width=14cm]{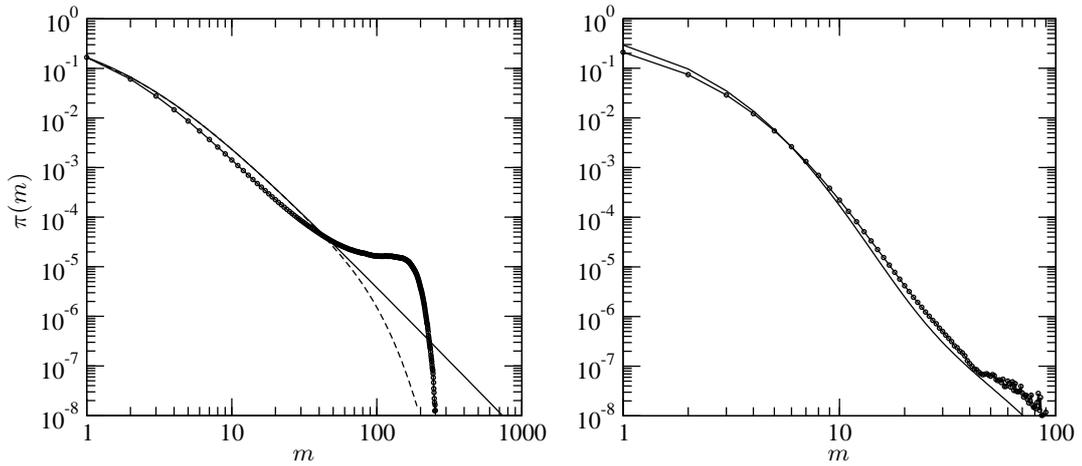}
\end{center}
\caption{\label{power-law+ER}Left: the distribution of balls for a network with degree distribution (\ref{piqgen}) with $b=3$, which gives $\rho_c=1$. Circles: computer simulation for $N=M=400$ and $\phi=30$. Solid line: asymptotic solution (\ref{pmb}) for $\phi\to\infty$, dashed line: the distribution for an infinite network but calculated for the cut-off $\phi=30$, from Eq.~(\ref{mom}) with Eq.~(\ref{piqgen}) inserted. Right: $\pi(m)$ for E-R graphs with $N=800,M=450,\bar{k}=8$ (circles). Solid line: theoretical solution, resembling a power-law but only in two decades of $m$. If continued for larger $m$, the curve deviates much from a straight line.}
\end{figure}

Coming back to the model (\ref{part3}) with weights given by
the moments of the degree distribution $\Pi(k)$, an interesting question emerges: can one
obtain a power-law distribution of balls by tuning $\Pi(k)$ so that the moments $\mu(m)\sim m^{-b}$, 
in order to mimic the critical behavior of a homogeneous system? The answer is in the affirmative. 
One can check \cite{zrp-short} that under the following choice:
\bq
\Pi(k) \propto (\phi-k)^{b-1}   \label{piqgen}
\eq
for $k<\phi$ and zero for $k>\phi$, the following
power-law in $\pi(m)$, 
\bq
\pi(m)\propto \Gamma(m+1)/\Gamma(m+b+1) \sim m^{-b}, \label{pmb}
\eq
is reproduced in the thermodynamic limit at the critical density 
$\rho_c=1/(b-2)$. The value $\phi$ plays a role of the maximal 
degree in the network. It depends approximately linearly on the average degree: $\phi\approx (b+1)\bar{k}$. 
Strictly speaking, Eq.~(\ref{pmb}) is valid only for infinite networks, 
for which $\phi\to\infty$. For finite $\phi$, the distribution $\pi(m)$ 
has a cut-off which scales linearly with $\phi$.
There is also another effect which disturbs the pure power law for finite networks. The power law in $\pi(m)$ comes from a superposition of exponential decays for nodes of different degrees: $\pi_i(m)\sim (k_i/\phi)^m$ for all nodes except the one with largest degree. On the most inhomogeneous node the condensation occurs similarly like in the inhomogeneous networks from the previous section. Since it takes some number of balls, it raises the critical density $\rho_c$ for which the power law is observed, and leads to the appearance of an additional peak in $\pi(m)$. On the left-hand side of Fig.~\ref{power-law+ER}, we show this effect for $N=M=400$ and $\phi=30$. In Ref.~\cite{zrp-short} it is shown that these finite-size effects becomes less and less important for larger networks.

In the above, we were interested in the question of how to tune
the degree distribution $\Pi(k)$ of the random network to obtain a desired
distribution $\pi(m)$ of particles in the FZRP on this network.
Actually, a simpler and more natural one is the opposite question:
can we determine the distribution of particles of the FZRP
on a random network with a given node degree distribution?
The answer is again positive. As an example let us
discuss what happens for some popular network ensembles, 
i.e. for random trees \cite{bbjk} and Erd\"{o}s-R\'{e}nyi (E-R) 
random graphs \cite{homnasz}. Both of them reveal a faster-than-exponential 
decay of the degree distribution: $\Pi(k)=e^{-1}/(k-1)!$ for trees 
and $\Pi(k)=e^{-\bar{k}}\bar{k}^k/k!$ for E-R graphs, and are effectively 
uncorrelated even for small networks, thus our method should perform well. 
On the right-hand side of Fig.~\ref{power-law+ER} we see results of numerical experiments. A quick glance on the figure would suggest that there is again a power-law behavior in $\pi(m)$. This is, however, not the case. 
By a simple analytic calculation one can show that one has to
do with a visual artifact caused by looking only on less than two decades in $m$. It can be estimated that the leading term in $\mu(m)$ is proportional to $\exp(m\log m)$ and grows for both the types of networks, so in the limit of $N\to\infty$ with a fixed density of particles there is always a condensation in these
models. For intermediate values of $M$, the dependence of the moments $\mu(m)$ on their order $m$ falls faster than exponentially but accidentally it resembles a quasi-power-law for small $m$. 

\section{MATTER-NETWORK INTERACTION} 
\label{sect:interaction}

Up to now we have considered the situation where the network had a fixed topology and the only dynamical part of the system were balls hopping between nodes. But in fact the model from the last section where we average over random networks,
can be regarded as a special kind of a dynamics of balls coupled to a very slowly rearranging network. The characteristic time scale $t_R$ for these rewirings is, however, much larger than the time scale $t_B$ for the evolution of balls, so from the ball's point of view the network is static or ``quenched''. One can pose the question, what changes when these two time scales become comparable. In this section we shall discuss this issue using as an example a slightly modified FZRP with
an explicit coupling between the two systems.

In this model particles propagate on a dynamically rearranging network.
At each time step of the process one of two changes is made:
either a particle is moved or the network is locally updated. It is realized
by a simple Monte Carlo process where we pick a node $i$ at random and
either with probability $1-r$ move  a ball like in Sec.~\ref{sect:statics}, 
or with probability $r$ we pick up a link $l$ at random and rewire one of its endpoints $j$ to $i$ with a Metropolis probability:
$\min\left\{ 1, w(m_i)/w(m_j)\right\}$. 
Here $w(m)$ is an arbitrary weight function which is an external
parameter of the model. The function $w(m)$ encodes the mutual 
coupling between the network topology and the particles.
If for example $w(m)$ decreases with $m$, the links tend to
avoid nodes with many balls while when it increases, 
the links tend to condense on such nodes.

The two characteristic time scales $t_B \propto 1/(1-r)$ 
and $t_R \propto 1/r$ can be tuned by the parameter $r$.
Thus changing $r$ we can explore different regimes of 
a matter-network interaction. The model has two obvious limits, 
for $r=0$ and $r=1$. For $r=0$ we get pure FZRP on a network 
fixed by the initial condition at the beginning of the process. 
In the limit of $r=1$ we have only rewirings and the distribution of balls is static. This is similar to hidden-variable models of the evolution of complex networks, see Refs.~\cite{hd1,hd2}. In particular, when $w(m)=const$ or the initial distribution of balls is homogeneous, the rewiring process produces the E-R graph because the rewiring is entirely random \cite{homnasz}.
The behavior of the model for $0<r<1$ strongly depends on the function $w(m)$. If $w(m)$ increases with $m$, even a small initial inhomogeneity in the degree distribution grows with time, because the inhomogeneity induces a condensation on irregular nodes, which is immediately amplified by the weight $w(m)$
which causes an attraction of new links to these nodes. This is a kind of avalanche reaction which leads to a condensation of balls and links on a single node.

If $w(m)$ is constant or decreases with $m$, then one can prevent
the condensation by increasing $r$. Because these two situations are similar to some extent, for simplicity we stick to $w(m)=1$. The network evolves now independently of balls. Because rewirings are random, the network topology quickly reaches a quasi-equilibrium state which is equivalent to E-R random graphs. The steady-state distribution of degrees $\Pi(k)$ is Poissonian. From the previous section we know that on the static network of that type, a condensation occurs on the node with the highest degree. Now, however, because the network is continuously rewired it may happen that the largest number of links hops from node to node. If this process is fast enough, one could expect that there is not enough time to form the condensate. For large $r$, any excess of balls appearing on a particular node could be quickly discharged and the balls became distributed quite uniformly in the network.

We will now present computer simulations. As we will see
the results confirm the phenomenological intuitive picture sketched above.
Moreover, we will be able to derive a characteristic
scaling of the parameter $r$ which balances the relative time
scales of the rewirings and FZRP dynamics and keeps the system
on the coexistence line between the condensed and the fluid phase.

\begin{figure}
\begin{center}
\psfrag{xx}{$m$} \psfrag{yy}{$\pi(m)$} \psfrag{x1}{$M$} \psfrag{y1}{$N\rho_c $} \psfrag{x2}{$m$} \psfrag{y2}{$\pi(m)$}
\includegraphics*[width=14cm]{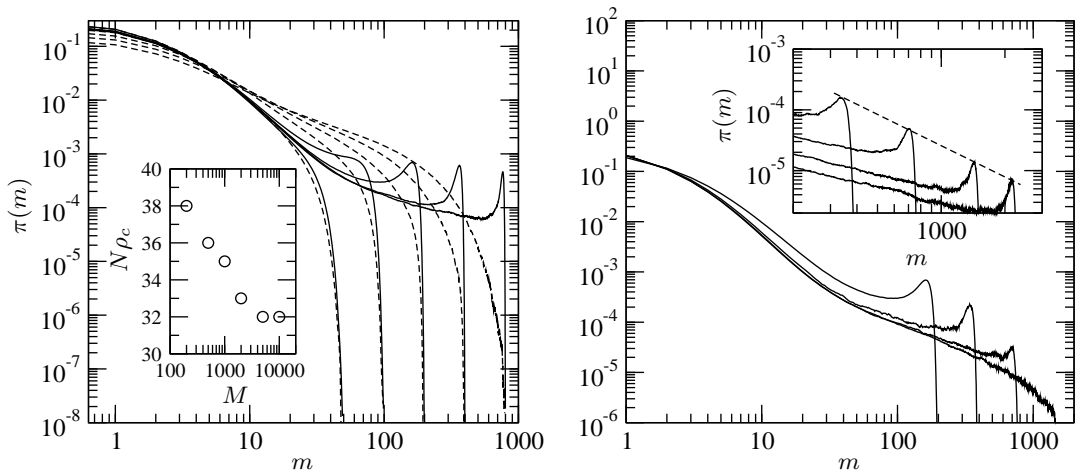}
\end{center}
\caption{\label{flor-n+m1}Left: plots of the occupation distribution $\pi(m)$ for fixed number of nodes and links $N=20,L=80$, different values of $M=50,100,200,400,800$ (curves from left to right) and two different $r=0.00005$ (solid lines) and $r=0.01$ (dashed lines). Inset: the difference between $M$ and the position of the peak in $\pi(m)$, for various $M$. This difference is by definition equal to $N\rho_c$ which allows for estimating $\rho_c\approx 1.6$ in the limit $M\to\infty$. Right: plots of $\pi(m)$ for fixed $r=0.00005, \rho=10$, average degree $\bar{k}=8$ and various $N=20,40,80,160$. The condensation disappears for large networks. Inset: if $r$ scales like $R/N^2$, the condensation is present also for large networks. All peaks for $N=40,80,160,240$ lie on the same line, $R=0.16, \rho=10, \bar{k}=8$.}
\end{figure}

\begin{figure}
\begin{center}
\psfrag{xx}{$m$} \psfrag{yy}{$\pi(m)$} \psfrag{x1}{$R$} \psfrag{y1}{$\rho $} \psfrag{f}{fluid} \psfrag{c}{condensed}
\includegraphics*[width=12cm]{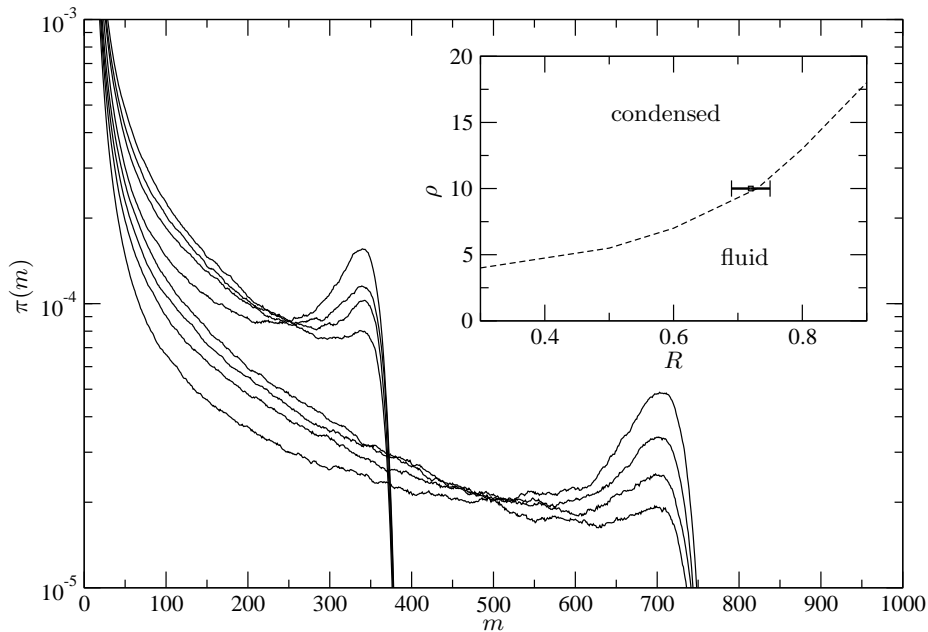}
\end{center}
\caption{\label{flor-r0}The distribution of balls for $R=0.16,0.32,0.48,0.64$. Left curves: $N=40$, right curves: $N=80$.
In both cases $\bar{k}=8,\rho=10$. Inset: the point represents an estimated value of $\rho_c(R_c)$ for $R_c\approx 0.72$. The dashed curve is a sketch
of the critical line separating the liquid and condensed phase.
}
\end{figure}

In Fig.~\ref{flor-n+m1}, left, we show plots of $\pi(m)$ for fixed network 
size $N$, for various rewiring probabilities $r$ and number of balls $M$.
We have chosen a relatively large value of the average degree 
$\bar{k}=2L/N=8$ to prevent the graph from splitting into 
disconnected parts during the rewiring. Let us first focus 
on the dependence of $\pi(m)$ on $M$ for $N,r$ fixed. When $r$ is small, there exists some critical density of balls $\rho_c$ above which the condensate appears in the system. The value of $\rho_c$ can be estimated assuming that the position of the peak in $\pi(m)$ is given by $\Delta M=M-N\rho_c$. This gives $\rho_c\approx 1.6$ for the network from Fig.~\ref{flor-n+m1}. When $r$ is large enough, the condensate is absent, regardless of the density of balls. The distribution $\pi(m)$ becomes more flat for increasing $M$ but it does not exhibit a peak.
It is obvious, that there must be some critical value $r_c$ which separates those two regimes. This critical point is, strictly speaking, determined by the ratio $t_R/t_B$ of the two time scales, and thus depends on the size of the system because the characteristic time for condensate formation depends on $N$. The rewiring rate $r$ must hence be properly rescaled to get a value independent of $N$.
On the right-hand side of Fig.~\ref{flor-n+m1} we see what happens when $r$ is fixed while increasing the size of the system, keeping the average degree and density of balls constant. The condensate, present for small networks, disappears for large $N$. But if we rescale $r$ with the system size $N$ as follows,
\bq 
r = R/N^2, 
\eq
where $R$ is some constant, then the peak stays in $\pi(m)$ even for large $N$. Moreover, the straight line on the log-log scale, going through the tops of the peaks for various $N$ corresponds to a scaling with $1/N$ of the peaks' area, convincing us that the scaling $\propto 1/N^2$ indeed indicates one node with the condensate.
In Fig.~\ref{flor-r0} we show plots of $\pi(m)$, with $r=R/N^2$, for different values of $R$. The estimated position of $R_c=N^2 r_c \approx 0.72(3)$ is the same for two different sizes $N=40,80$ and $\bar{k}=8, \rho=10$. This confirms again that the chosen scaling is correct.
Repeating this procedure, one could obtain a phase diagram $\rho(R)$. A sketch of the diagram is shown in the inset of Fig.~\ref{flor-r0}. The critical line $\rho_c(R)$ separates two phases: the liquid one for $\rho<\rho_c$, and the condensed one for $\rho>\rho_c$. The point represents the result of simulations. We know also that $\rho_c(N^2)=\infty$ since in this case the system is homogeneous from the balls' point of view and no condensation is possible. For $R=0$, the critical value of $\rho$ depends on the average degree, but in general is non-zero (see plots for E-R graphs from Sec. \ref{sect:fluct}).

\section{SUMMARY}

Despite its simplicity, free zero-range processes on networks exhibit a very rich
behavior, being very well suited to address complicated
questions concerning dynamics on networks including 
non-equilibrium effects or interactions between network topology
and dynamics of degrees of freedom defined on the network. 
Such questions are important for a deeper understanding 
of the relation of the structure and functionality of complex systems. 
Some of them were explicitly discussed in this paper but some others
like a systematic study of the back-reaction of matter and network is still open. %for the future. 
Surprisingly,
similar questions have already been addressed in a completely 
different field of research, namely in the field of quantum gravity
where random graphs (dynamical triangulations)
were used to approximate Riemannian manifolds (Euclidean version
of space-time). Such Riemannian manifolds were let to interact 
with matter fields defined on space-time. It turns out that 
the dynamical geometry of such manifolds 
affects critical properties of the models defined on them.
For example, the Ising model on two-dimensional random triangulation
is known to have different critical indices than its counterpart on
a single two-dimensional graph which has the Onsager exponents.
However, one of the most surprising effects observed in the Ising
model on dynamical triangulation
is that not only randomness of the graph influences the matter
field but that there is a strong backreaction from the behavior
of the matter field on the geometry of the dynamical triangulation.
When the matter field becomes critical it strongly influences 
geometrical properties of the underlying triangulation. 
Certainly similar effects should also be observed 
in complex systems where the behavior of degrees of freedom in
the system may act collectively to change its skeleton and the
underlying network structure on which the signals propagate.
We hope to address this issue in the future.

\section*{Acknowledgments}

This work was supported in part by the EC-RTN Network ``ENRAGE'' under 
grant No. MRTN-CT-2004-005616, an Institute Partnership grant 
of the Alexander von Humboldt Foundation,
an EC Marie Curie Host Development Fellowship under 
Grant No. IHP-HPMD-CT-2001-00108 (L.~B. and W.~J.), an EC ToK 
Grant MTKD-CT-2004-517186 ``COCOS'' and a Polish Ministry of Science
and Information Society Technologies Grant 1P03B-04029 (2005-2008)
(Z.~B.).
B.~W. thanks the German Academic Exchange Service (DAAD) for support.

\end{document}